\documentclass[%
 reprint,
 amsmath,amssymb,
 aps,
prb,
]{revtex4-2}
\usepackage{physics}
\usepackage{graphicx}
\usepackage{dcolumn}
\usepackage{bm}
\usepackage{xcolor}
\usepackage{mathtools,leftindex,tensor,mhchem}

\begin{document}


\title{Measurement Induced Asymmetric Entanglement in a Deconfined Quantum Critical Ground State}

\author{K.G.S.H. Gunawardana}
\email{harshakgs@gmail.com, korala.gunawardana@tuni.fi}
 \affiliation{Computational Physics Laboratory, Physics Unit, Faculty of Engineering and Natural Sciences, Tampere University,P.O.Box 692, FI-33014 Tampere, Finland.}




\date{\today}

\begin{abstract}
In this work, we numerically study the effect of weak measurement on deconfined quantum critical point(DQCP).   Particularly, we consider the ground state of an one-dimensional spin $1/2$ system with next-nearest-neighbour exchange interactions($K$), which shows analogues phase transition to DQCP in the thermodynamic limit. This system is in the ferromagnetic phase below the critical exchange interaction $K_c$ and in the valance bond solid phase above $K_c$.  The weak measurement is carried out by coupling a secondary ancilla system to the critical system via unitary interactions and later measuring the ancilla spins projectively. 
We numerically calculate entanglement entropy,correlation length, and order parameters of leading post-measurement states using uniform matrix product state representation of the quantum many-body state in the thermodynamic limit. 
We report  asymmetric restructuring of entanglement of the post measurement states across the phase boundary under weak measurements.  
Especially, the trajectory $\left(\downarrow \downarrow\right)$ describing a uniform measurement outcome given the all ancilla spins initiated in the same $\left(\downarrow \right)$ state, shows anomalous entanglement when increasing the strength of weak measurement. The bipartite entanglement entropy strongly increases  when $K<K_c$ whereas it weakly decreases when $K>K_c$.  
We argue with numerical evidences that observed asymmetry in entanglement would lead to a weak first order phase boundary in the thermodynamic limit. We also discuss important aspects in experimental observation of measurement induced effects linked to the   strength of weak measurement  and probability of post-measurement states.   

\end{abstract}

\maketitle


\section{\label{sec:intro}Introduction} 
The deconfined quantum critical point(DQCP) describes a continuous phase transition between phases that spontaneously break distinct symmetries which preclude in the conventional Landau-Ginzburg-Wilson framework \citep{sentil2004science}. This phenomena could be explained by the existence of deconfined fractional  degrees of freedom  coupled to emergent gauge fields at the critical point \citep{sentil2004prb1,sentil2004prb2,sentil2006prb,didier2006prb,nahum2015prl,shao2016science,wang2017prx,yan2017prx, nvsen2018prb,senthil2023review, lei2024prb}.
The DQCP has been studied extensively in recent years with rigorous theoretical frameworks and extensive numerical calculations\citep{ kaul2012prl, pujari2013prl, yang2014prb, ling2016prb, dimitri2017prb, francesco2018prb, jian2018prb, zhang2018prl, lee2019prx, jiang2019prb, roberts2019prb, mudry2019prb, huang2020prr, zhao2020prl, leyna2021prb, yang2021pre, roberts2021prb, xi2022cps, li2023prb, liu2023prl, lee2023prl, xi2023prb, zhang2023prl, romen2024spst, liu2024prb, wang2025prb, cui2025review}. 
However, the experimental realization of this exotic quantum phase transition has been a challenging task so far \cite{cui2023science,guo2025comphys}.  In this work, we investigate a different aspect arisen when the deconfined quantum critical system is exposed to an external environment and being probed or  monitored by a measurement apparatus. 
    
In general, decoherence may occur when a quantum system interact with its environment or been measured coupling to a measurement apparatus\cite{zurek2003RevModPhys}. 
The initial pure state of a system would decohere into a mixed state due to interaction with the extra degrees of freedom of the measurement apparatus. The problem arise when it is unable to account for all the measurement outcomes and hence the mixed state  is collapsed into a subset losing information of the initial state. In such cases, the physics of individual post-measurement states  carries vital information of quantum processes which depend not only on the initial state but also the nature of the system-measurement apparatus coupling and measurement outcomes.  
This phenomena has been received much attention in recent years in relation to the measurement induced entanglement phase transition of which the measurements often act as non-unitary processes limiting the amount of entanglement in the system\cite{li2018prb,skinner2019prx,cao2019scipost,fuji2020prb,bao2020prb,turkeshi2020prb,takaaki2022prl,ali2023prl, sang2023prxq,fisher2023review,cecile2024prr,igor2025prb}. The local measurements would restructure the entanglement non-locally, particularly in the measurement induced entanglement(MIE), a phenomena received much attention recently in theory and numerical simulation, the entanglement would also enhance \cite{lin2023quantum, zihan2024prb, zhang2024pra, paviglianiti2024quantum, avakian2025prr}. 
On quantum critical ground states, it has been reported that measurement would results highly non-trivial effects on correlations and entanglement depending on the measurement basis and the outcome.   
The quantum critical systems studied include conventional quantum critical phases such as Luttinger liquid \cite{garratt2023prx, ashida2024prb,tang2024arxiv}, transverse field Ising model\cite{yang2023prb,zack2023prb,sara2023prx,sala2024prx} and $(2+1)$-dimensional quantum critical points\cite{lee2023prxq}. 

In this work, we study the effect of weak measurement on the deconfined quantum critical point by numerically calculating the entanglement entropy, correlation length and order parameters of leading post-measurement states  across the phase boundary. 
A factual approach is implemented considering the critical system coupled to an alcilla system via unitary interactions and subsequently measuring the ancilla spins projectively. \cite{janet2010pra, timothy2013pra, sara2023prx}.  It is considered a one dimensional system of spin $1/2$ moments with \textcolor{black}{next-nearest-neighbour exchange} interaction($K$) whose  ground state shows many analogies to the DQCP \cite{roberts2019prb, jiang2019prb}. 
The system is in ferromagnetic(zFM) phase below a critical coupling $K_c$ and in a valance bond solid phase above(VBS) $K_c$. We derive two type of measurement operators: $X(u,\alpha)$ and $Z(u,\alpha)$, correspond to two form of critical-ancilla unitary interactions( See eq. (\ref{eq:Uxx}) and (\ref{eq:Uzx})). 
In our formalism, the action of measurement operators is a combination of unitary operations ($u$) and non-unitary actions($\alpha$) contributing to restructure the entanglement across the system. When the parameter $\alpha =0$, the measurement operators act as local unitary operators. 
In this context, the parameter $u$ govern the Born-rule probability of a measurement outcome in the weak measurement limit which is defined as $\alpha<<1$. 

\begin{figure}[ht]
\centerline{\includegraphics[scale=0.38]{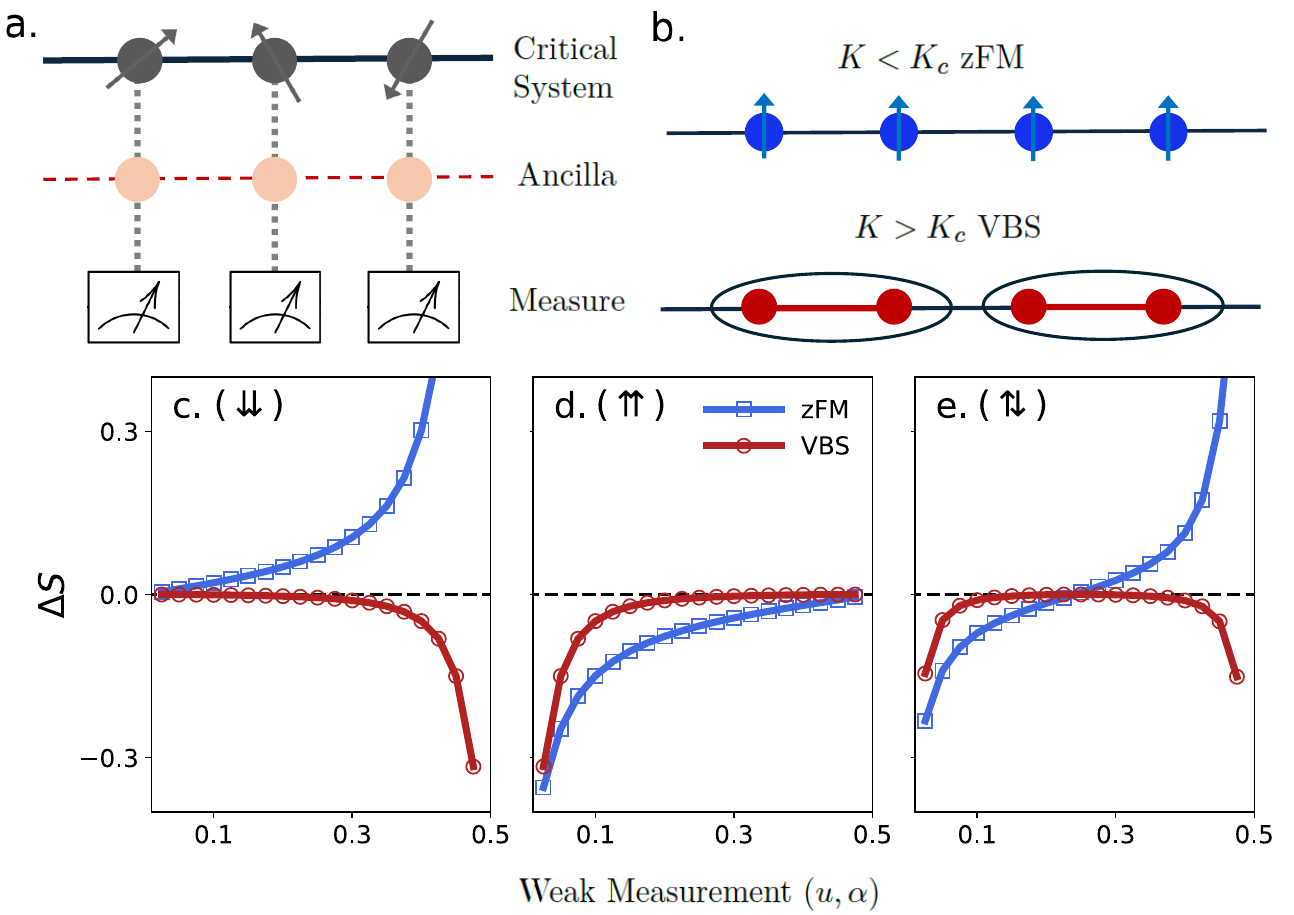}}
\caption{The panel a. shows a schematic of the infinitely long critical system coupled to ancilla spins and the ancilla spins are measured projectively with a measurement apparatus. The panel b. depicts the zFM phase below the critical coupling $K<K_c$ and the paired spins in singlet VBS phase(red solid circles connected with solid red line) above the critical coupling $K>K_c$. The lower panels(c. d. and e.) depict $\Delta S$, the shift of bipartite entanglement entropy of post-measurement states from that of critical ground state for 3 leading measurement outcomes, against the unitary parameter $u$ at $\alpha=0.004$. The measurement outcomes $\left(\downarrow \downarrow \right)$, $\left(\uparrow \uparrow \right)$, and $\left(\uparrow \downarrow \right)$ denote the periodic extension of the spin states($z$-basis) in side the parenthesis  in an infinitely long chain. The blue solid lines with squares represent the $\Delta S$ of ferromagnetic(zFM) phase when $K<K_c$. The red solid lines with circles represent the $\Delta S$ of valance bond solid (VBS) phase when $K>K_c$. Note the asymmetric development of $\Delta S$ between two phases under the weak measurement.}
\label{fig.1}
\end{figure}

We find an asymmetric restructuring of bipartite entanglement entropy($S$) in the post-measurement states with the $Z$ type of weak measurement as shown in the lower panels of Fig.\ref{fig.1}. It is noted  the highly asymmetric entanglement across the phase boundary for three different measurement outcomes. 
The measurement outcomes $\left(\downarrow \downarrow\right)$, $\left(\uparrow \uparrow\right)$ and $\left(\uparrow \downarrow\right)$ represent the periodic extension of the spin states inside the parenthesis on an infinitely long ancilla system initiated in the $\ket{\downarrow}$ state. 
It is evidenced that the observed asymmetry in  $S$ is directly related to the restructuring of local correlations in the system under the weak measurement. 
For instance, the enhancement of $S$ in the measurement outcome $\left(\downarrow \downarrow\right)$ when $K<K_c$ could be a result of converting short-range correlations to long-range correlations. We numerically verify that the correlation length $\xi$ increases drastically when $K<K_c$.  On the other hand, $\xi$ decreases in the region $K>K_c$ which is consistent with the observed reduction in $S$ under weak measurements. Further, we observe that above asymmetric entanglement leads to a large gap in the correlation length, $\Delta \xi$,  at the  critical point $K_c$ which grows monotonically with the MPS bond dimension $\chi$. We argue that the developing $\Delta \xi$ in the vicinity of  $K_c$ of the post-measurement states would lead to a weak first order transition in the thermodynamic limit. 
 
In proceeding, this paper is organized as follows. First it is given a detail description of the critical system and the measurement protocol in section \ref{sec:Model}.  In section \ref{sec:Results}, first, it is presented the trivial results when the parameter  controlling the weak measurement strength $\alpha=0$. Next, we present the numerical results of entanglement entropy and the correlation length of leading post-measurement states under weak measurement and discuss the nature of the zFM to VBS phase transition. Finally, we discuss important aspects on measurement probability and strength of weak measurement  in relation to the experimental realization of measurement induced phenomena. A concluding discussion is presented in the section \ref{con:disc}. 

\section{\label{sec:Model}The Model and Measurement Protocol}
\subsection{Model}
The Hamiltonian of the one-dimensional system of spin $1/2$ moments is given below. 
\begin{multline}
H=\sum_j\left(-J_x\sigma^x_j\sigma^x_{j+1} - J_z\sigma^z_j\sigma^z_{j+1}\right) \\  + K \left(\sigma^x_j\sigma^x_{j+2} + \sigma^z_j\sigma^z_{j+2}\right), 
\label{Hqs}
\end{multline}
The parameters $J_x$ and $J_z$ represent the nearest-neighbour exchange interactions and the parameter $K$ represents the isotropic next-nearest-neighbour exchange interaction. The Pauli spin operators act on site $j$ is represented by $\sigma^{x,z}_j$.  Recent numerical studies show that the ground state of the above Hamiltonian has many analogies to the  two dimensional Deconfined Quantum Critical phase transition \cite{roberts2019prb, jiang2019prb}. 
Particularly, at anisotropic nearest-neighbour exchange interactions which breaks $U(1)$ symmetry, the system shows ferromagnet  to valance bond solid phase transition as varying the long range exchange interaction $K$. 
At low values of $K$ the spins are ordered in $\sigma^z$ direction to form ferromagnetic phase (zFM) and at higher values of $K$ the spins form singlets resulting an ordered pattern of valance bonds in the valance bond solid(VBS) phase (See Fig.\ref{fig.1} upper panel right). 
The study of the ground state of the above Hamiltonian using variational uniform matrix product states (VUMPS) algorithm and the finite size entanglement scaling with MPS bond dimension ($\chi$) proves the existence of a continuous  phase transition at the thermodynamic limit. 
We follow similar approach to construct the DQCP ground state using the VUMPS algorithm \cite{vumps_zauner2018prb, vumps2_zauner2018prb, Vanderstraeten_2019}.   
Thus, it allows to study effect of measurement in the thermodynamic limit by scaling with the bond dimension $(\chi)$.

\subsection{Measurement Protocol}
  
Figure \ref{fig.1} (top panel) shows a schematic of the infinitely long critical system coupled to ancilla of spin 1/2 moments.  Each spin in the critical system is coupled to an spin of the ancilla via unitary coupling $U_j$. We consider two form of unitary coupling represented by $U_{\sigma_x \otimes \widetilde{\sigma_x}}(u,\alpha)$ and $U_{\sigma_z \otimes \widetilde{\sigma_x}}(u,\alpha)$ as given by the equations (\ref{eq:Uxx}) and (\ref{eq:Uzx}). 
Following the unitary coupling, each ancilla spin of the composite system is measured projectively. In this section, we briefly discuss the construction of set of measurement operators for above process introducing the notations used in this paper. 

The initial state of the critical system and the ancilla before coupling can be written as,
\begin{equation}
\ket{\psi_i}=\ket{\psi_{gs}} \ket{\psi_a}, 
\end{equation}
where $\ket{\psi_{gs}}$ is the critical ground state of the Hamiltonian (\ref{Hqs}) and $\ket{\psi_a}$ is the initial state of the ancilla. 
The quantum state of the composite system after the unitary coupling can be expressed as , 
\begin{equation}
\ket{\psi_U}=\prod_j U_j \ket{\psi_{gs}} \ket{\psi_a},
\end{equation}
where the unitary operator $U_j$ couples the $j^{th}$ spin of the critical system to the $j^{th}$ spin of the ancilla. 
Next the ancilla spins are measured projectively with the projector $\ket{\psi_{a'}}\bra{\psi_{a'}}$.  The state of the critical system after the projective measurement can be expressed as, 
\begin{equation}
\ket{\psi_{m}}=\frac{1}{\sqrt{p_m}}\bra{\psi_{a'}}\prod_j U_j \ket{\psi_{gs}} \ket{\psi_a},
\end{equation}
where $\ket{\psi_{m}}$ is the post measurement state of the critical system and $p_m=\bra{\psi_m} \ket{\psi_m}$ is the probability of the post-measurement state $\ket{\psi_m}$.  
Assuming the ancilla is initially in a product state, the post measurement state can be written as, 
\begin{equation}
\ket{\psi_{m}}=\frac{1}{\sqrt{p_m}}\left(\prod_j \bra{a'_j} U_j \ket{a_j}\right) \ket{\psi_{gs}},
\end{equation}
where $\ket{a_j}$ and $\ket{a'_j}$ are initial state and the measurement outcome of the ancilla spin at site $j$ respectively. 
Now the measurement operators (Kraus operators) acting on the site $j$ of the critical ground state can be defined as, 
\begin{equation}
M_j^{a'a}=\bra{a'_j} U_j \ket{a_j}.
\label{eq:M_op}
\end{equation} 
For a given initial state of the $\ket{a}$ the $M_j^{a'a}$ satisfy the condition for a positive operator-valued measure(POVM),
\begin{equation}
 \sum_{\{a'\}}\left(M^{a'a}_j\right)^{\dagger} M^{a'a}_j=1, 
 \label{eq:povm}
\end{equation} 
which ensure the normalization of the total Born-rule probability. 

The ground state of the critical system is obtained in the uniform matrix product state (uniform MPS) representation using the VUMPS algorithm. It is considered two spin moments in the unit cell to accommodate the VBS phase in the uniform MPS ground state.
We consider that the ancilla is also following the same periodicity as the critical system for the simplicity of numerical implementation in the uniform MPS representation. In this work, we only consider the $z$-basis measurement and the spin states $\ket{\uparrow}$ and $\ket{\downarrow}$ are the eigenstates of the Pauli operator $\sigma^z$. 
All the ancilla spins are initiated in the $\ket{\downarrow}$ state. In the basis of two spins in the unit cell, the measurement outcome $\left(\downarrow\downarrow\right)$ represent a uniform trajectory measuring all the ancilla spins in $\ket{\downarrow}$ state. Similarly, the measurement outcome $\left(\uparrow\downarrow\right)$ represent a trajectory measuring alternative spin orientation in the infinitely long ancilla.  
Thus, the spin states inside the parenthesis represent periodic repetition over an infinitely long system in the numerical calculation. In general, this procedure can be extended to larger unit cells and measurement outcomes such as $\left(\uparrow \uparrow \downarrow \cdots \uparrow \right)$ can be considered. However, we confirm that such measurement outcomes does not produce qualitatively different results reporting in this analysis. 

Consequently, the post measurement states correspond to the  measurement outcome $\left(\downarrow \downarrow \right)$ is written as,  
\begin{equation}
\ket{\psi_{\left(\downarrow \downarrow \right)}}= \prod_k \left(M_1^{\downarrow \downarrow}M_2^{\downarrow \downarrow}\right)_k \ket{\psi_{gs}},
\label{eq:PMstate1}
\end{equation}
and that of measurement outcome $\left(\uparrow \downarrow \right)$ is written as,
\begin{equation}
\ket{\psi_{\left(\uparrow \downarrow \right)}}= \prod_k \left(M_1^{\uparrow \downarrow}M_2^{\downarrow \downarrow}\right)_k \ket{\psi_{gs}} ,
\label{eq:PMstate2}
\end{equation}
where the index $k$ refers to the unit cell of the critical chain.
The measurement operators $M_1^{\uparrow \downarrow}$ and $M_2^{\downarrow \downarrow}$  are acting on the $1^{st}$ and the $2^{nd}$ spin moments in the unit cell of the critical system respectively. 
\\
\subsection{Unitary coupling and measurement operators}
The form of the unitary coupling between a pair of spin moments in the critical system and the ancilla can be  defined as below. 
\begin{equation}
 U_{\sigma_x \otimes \widetilde{\sigma_x}}(u,\alpha)=e^{-i \left(u \sigma_{x}-\alpha \mathbb{I} \right)\otimes \widetilde{\sigma_{x}}},
\label{eq:Uxx}
\end{equation}
and 
\begin{equation}
 U_{\sigma_z \otimes \widetilde{\sigma_x}}(u,\alpha)=e^{-i \left(u \sigma_{z}-\alpha \mathbb{I} \right)\otimes \widetilde{\sigma_{x}}}.
\label{eq:Uzx}
\end{equation}
It is assumed that the parameters $u$ and the $\alpha$ are spatially independent. The Pauli spin operators with sign  $\widetilde{\sigma}$  act on the ancilla spins while the others act on the critical spins.  The $\mathbb{I}$ represent the $2\times 2$ identity matrix. 
Following the eq.(\ref{eq:M_op}), we derive two sets of measurement operators $X$ and $Z$ correspond to the unitary coupling given in equation (\ref{eq:Uxx}) and (\ref{eq:Uzx}) respectively. 
\begin{align}
X^{\downarrow \downarrow}(u,\alpha)=\begin{pmatrix}
\cos(u)\cos(\alpha) & \sin(u)\sin(\alpha ) \\
\sin(u)\sin(\alpha) & \cos(u)\cos(\alpha )
\end{pmatrix}
\label{eq:Xdd}
\end{align}
\begin{align}
X^{\uparrow \downarrow}(u,\alpha)=\begin{pmatrix}
\cos(u)\sin(\alpha ) & -\sin(u)\cos(\alpha ) \\
-\sin(u)\cos(\alpha ) & \cos(u)\sin(\alpha )
\end{pmatrix}
\label{eq:Xud}
\end{align}
and
\begin{align}
Z^{\downarrow \downarrow}(u,\alpha)=\begin{pmatrix}
\cos (u+\alpha) & 0 \\
0 & \cos (u-\alpha)
\end{pmatrix}
\label{eq:Zdd}
\end{align}
\begin{align}
Z^{\uparrow \downarrow}(u,\alpha)=\begin{pmatrix}
 \sin (u+\alpha) & 0 \\
0 & -\sin (u-\alpha)
\end{pmatrix}
\label{eq:Zud}
\end{align}
It should be noted that the the above sets of measurement operators reduce to unitary form when $\alpha=0$ yielding, $\left(X^{\downarrow \downarrow}\right)^{\dagger}X^{\downarrow \downarrow}=p^2\mathbb{I}$ and $\left(X^{\uparrow \downarrow}\right)^{\dagger}X^{\uparrow \downarrow}=q^2\mathbb{I}$  satisfying $p^2+q^2=1$. The $Z$ operators also satisfy the   above unitary condition at $\alpha=0$.  Thus, the measurement operators $X(u,0)$ and $Z(u,0)$ act as local unitary operators.
In this limit these operators are incapable of restructuring the entanglement across the critical system. When $\alpha \neq 0$, the strength of the measurement ($\lambda$) can be quantified as a linear function of the matrix elements contributing to the non-unitary nature of the measurement operators. For $X$ type measurement operators, the strength of the measurement for the operator in eq.(\ref{eq:Xdd}) can be given by the off-diagonal matrix element, $\lambda=\sin(u) \sin(\alpha)$  and for the operator in eq.(\ref{eq:Xud}) it can be given by the diagonal matrix element $\lambda=\cos(u) \sin(\alpha)$.  For $Z$ type measurement operators, it is the difference of the magnitudes of the diagonal elements. Thus, the $\lambda=\cos(u+\alpha)-\cos(u-\alpha)$ and $\lambda=\sin(u+\alpha)-\sin(u-\alpha)$ for the operators given in eq.(\ref{eq:Zdd}) and (\ref{eq:Zud}) respectively. 
 We consider a weak limit where the $\alpha<<1$.
It can be shown that for $\alpha<<1$, the strength of the weak measurement, $\lambda \sim \alpha \sin(u)$ and $\lambda \sim \alpha \cos(u)$ for measuring an alcilla spin in $\bra{\downarrow}$ state and $\bra{\uparrow}$  state  respectively for both type of measurement operators given the ancilla spin is initiated in the $\ket{\downarrow}$ state. Thus, the $\alpha$ is a measure of maximum strength of the weak measurement as varying the unitary parameter $u$. 

In proceeding, the post-measurement states given in equation (\ref{eq:PMstate1}) and (\ref{eq:PMstate2}) are  labelled as $\ket{\psi_{\left(\downarrow \downarrow \right)}^{X/Z}}$ and  $\ket{\psi_{\left(\downarrow \uparrow \right)}^{X/Z}}$ respectively to indicate the type of measurement operators implemented. 
The observables are calculated in the corresponding post-measurement state of a given measurement outcome.

\section{\label{sec:Results}Results}

\subsection{Non-entangling limit, $\alpha =0$. }
\begin{center}
\begin{figure}[ht]
\includegraphics[scale=0.33]{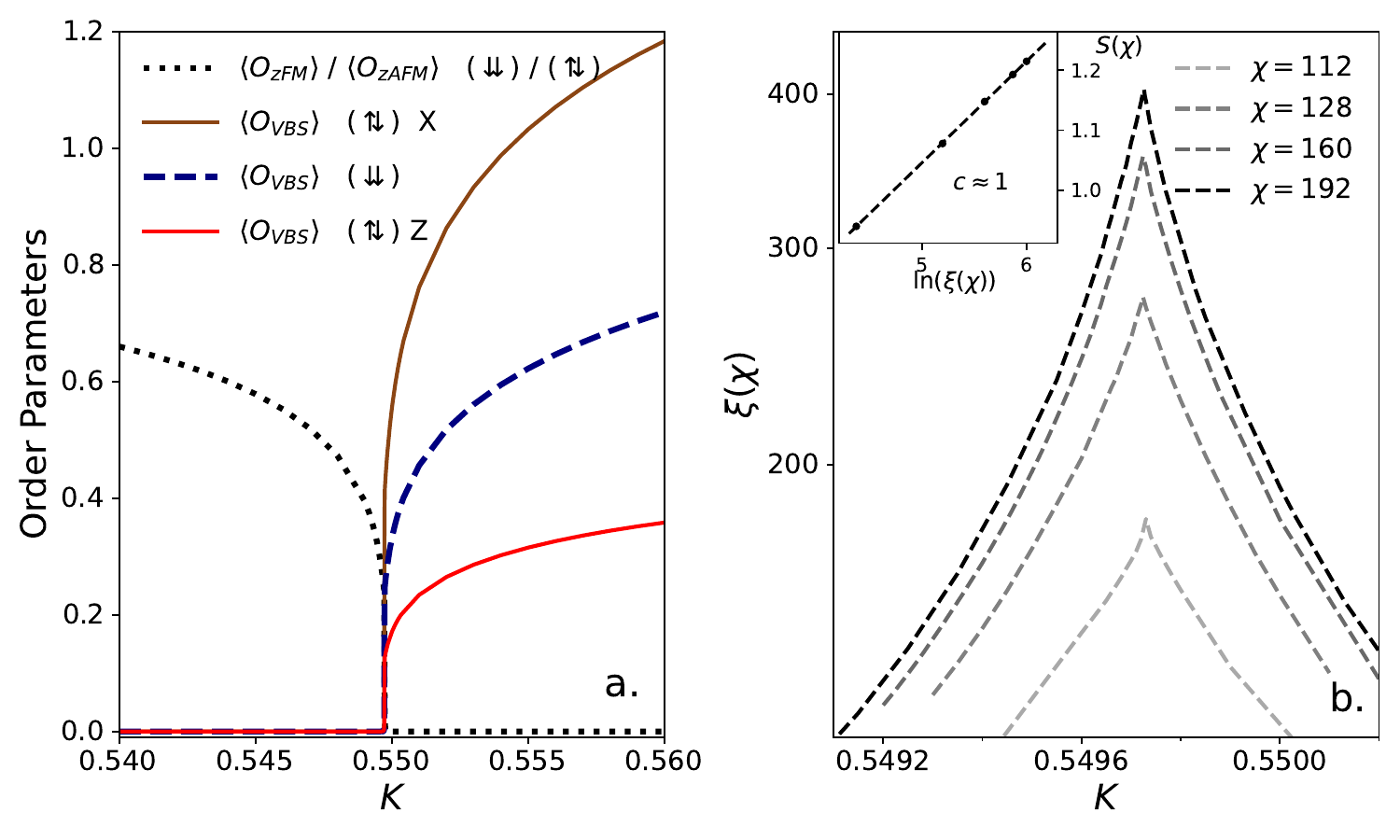}
\caption{\label{fig:2} a.) Numerically calculated order parameters as a function of the long range exchange interaction $K$. The results are calculated at $\chi=192$ using VUMPS algorithm. The dotted and the dashed lines represent the  $\langle O_{zFM}\rangle $  and $\langle O_{VBS}\rangle $  of the ground state of the critical system. The post-measurement states are calculated at $\alpha=0$ and $u=1/10$. The order parameters of the measurement outcome $\left(\downarrow \downarrow\right)$ with $X$ and $Z$ type measurements share the same dotted and dashed lines. The $\langle O_{zFM}\rangle$ and $\langle O_{zAFM}\rangle$ for the measurement outcome $\left(\uparrow \downarrow \right)$ also share the same dotted line for $Z$ and $X$ type measurements respectively. The $\langle O_{VBS}\rangle $ for $X$ type measurement is represented by the solid brown line and the that of $Z$ type measurement is represented by the solid red line. b.) The numerically calculated correlation length $\xi(\chi) $ at different values of MPS bond dimension $\chi $. The ground state and the post-measurement states at $\alpha=0$ shares the same correlation length $\xi(\chi)$. The peak value of $\xi(\chi)$ at the vicinity of the critical point diverges in the limit $\chi \rightarrow \infty$. The inset shows the  entanglement entropy $S(\chi)$ versus the $\ln(\xi(\chi))$ yielding the slope $\approx 1/6$ with effective central charge $c\approx 1$.}
\end{figure}
\end{center}

The $\ket{\psi_{gs}}$ of the Hamiltonian (\ref{Hqs}) when the nearest neighbour exchange interactions $J_x=0.5$ and $J_z=1.5$  is calculated in the thermodynamic limit with finite bond dimension($\chi$) using the VUMPS algorithm. The maximum bond dimension employed in the work is $\chi=192$. In the uniform MPS representation the system is considered as a periodic system with $2$ spin moments in the unit cell. 
The post measurement state is obtained by applying the measurement operators periodically to the uniform MPS ground state as in eq. \ref{eq:PMstate1} and \ref{eq:PMstate2}. 

First, the quantum phases are analysed by calculating the order parameters of valance bond solid $\langle O_{VBS}\rangle$, ferromagnetic $\langle O_{zFM}\rangle$ and anti-ferromagnetic $\langle O_{zAFM}\rangle$ phases (refer to eq.\ref{eq:odp}). 
We consider the $U_{\sigma_x \otimes \widetilde{\sigma_x}}$ and $U_{\sigma_z \otimes \widetilde{\sigma_x}}$ unitary coupling in the non-entangling limit where $\alpha=0$. 
Fig.\ref{fig:2} a.) shows the calculated order parameters as a function of $K$. The dotted and dashed lines represent the order parameters of the zFM and the VBS phases of the critical ground state. 
There is a phase transition from zFM to VBS at a pseudo-critical coupling $K_c(\chi)$ which depends on the bond dimension ($\chi$).
Both order parameters vanish simultaneously as reaching the critical coupling $K_c$ in the limit of $\chi \rightarrow \infty $. The reader is directed to ref. \cite{roberts2019prb} for more details of the ctitical phase transition of this system. 

The numerical calculation produce the trivial results for post-measurement states at $\alpha=0$. 
The $\ket{\psi_{\left(\downarrow \downarrow \right)}^{X/Z}}$ shows the same zFM to VBS phase transition. The $\langle O_{zFM}\rangle$ and $\langle O_{VBS}\rangle$ shares the dotted and dashed lines in the Fig. \ref{fig:2} a.) respectively for the measurement outcome $\left(\downarrow \downarrow \right)$ with $X$ and $Z$ type of measurement operators.  
Below the pseudo-critical coupling $K_c(\chi)$,  $\ket{\psi_{\left(\downarrow \uparrow \right)}^{Z}}$ is also in zFM phase with the same amplitude of $\langle O_{zFM}\rangle$ represented by the dotted line.
However, in $\ket{\psi_{\left(\uparrow \downarrow \right)}^{X}}$, measurement flips the spin state at odd numbered sites and hence the the initial zFM phase is fully converted to  zAFM phase. 
Thus, the amplitude of $\langle zAFM \rangle$ also shares  the dotted line when $K<K_c(\chi)$. In the Appendix \ref{app:orderp}, we show that the order parameters of the post measurement states at $\alpha=0$ follow the same critical behaviour when $\chi \rightarrow \infty $.

Some insight on the above observations can be drawn by considering the Hamiltonian under above local unitary transformations. For $X$ type $(\uparrow \downarrow)$ measurements , the Pauli spin operators at odd positions($j$) of the Hamiltonian (\ref{Hqs}) transform as, $\sigma^x_{j} \rightarrow \sigma^x_{j}$ and $\sigma^{z,y}_{j} \rightarrow -\sigma^{z,y}_{j}$. As a result, the nearest neighbour interactions of the Hamiltonian favour anti-ferromagnetic ground state. 
For $Z$ type $(\uparrow \downarrow)$ measurements , the Pauli spin operators at odd positions($j$) of the Hamiltonian \ref{Hqs} transform as, $\sigma^x_{j} \rightarrow -\sigma^x_{j}$ and $\sigma^{z,y}_{j} \rightarrow \sigma^{z,y}_{j}$. With dominating ferromagnetic interactions $J_z>J_x$, the sign change of the transverse nearest neighbour interaction term has no effect on the ground state when $K<K_c(\chi)$.
In the region $K>K_c(\chi)$, $X$ and $Z$ type measurement operators result distinct effects on the order parameter $\langle O_{VBS}\rangle$. The $X$ type measurement operators favours the VBS phase and it is observed a significant increment of  $\langle O_{VBS}\rangle$ (solid red line in Fig. \ref{fig:2}), whereas $Z$ type measurement decreases the order parameter amplitude  $\langle O_{VBS}\rangle$ (solid brown line Fig. \ref{fig:2}a).  It must be noted that the above order parameter amplitudes are independent of the unitary parameter $u$. 

Next, we calculate the correlation length $\xi(\chi)$ of post measurement states. It is noted that the critical ground state and the post measurement states share the same $\xi(\chi)$ at $\alpha=0$. The $\xi(\chi)$ is calculated from the eigenvalue spectrum of the MPS transfer matrix. The correlation length is given by $\xi(\chi)=-n/\ln(\lambda_2/\lambda_1)$, where $n$ is the number of spin moments in the unit cell and,  $\lambda_1$ and $\lambda_2$ are first and second largest eigenvalues of the transfer matrix. The diverging correlation length as  $K$ approaching the true critical point $K_c$ is a characteristic feature of the continuous phase transition. Fig. \ref{fig:2}b.) shows $\xi(\chi)$ as a function of $K$ across the critical coupling for different values of $\chi$.   The peak values of the $\xi(\chi)$ of the zFM (also zAFM)  and the VBS phases occur at the same pseudo-critical coupling. Moreover, the both phases have the same peak value of $\xi(\chi)$ at the pseudo-critical coupling and diverge in the limit $\chi\rightarrow \infty$,  which is a characteristic of a continuous phase transition.

The von Neumann entanglement entropy is calculated using singular value decomposition(SVD) of approximated finite size MPS of the uniform MPS. 
We calculate specially averaged von Neumann entanglement entropy, denoted by $S(\chi)$ by employing a bipartite cut at the first and the second bond of the unit cell. The $S(\chi)$ near the vicinity of the critical point follows logarithmic scaling, $S(\chi)=\frac{c}{6}\ln(\xi(\chi))$ given the central charge $c\approx 1$ as shown in the inset of the Fig.\ref{fig:2}b.).

\subsection{Weak measurement, $\alpha << 1$. }
\begin{center}
\begin{figure}[ht]
\includegraphics[scale=0.32]{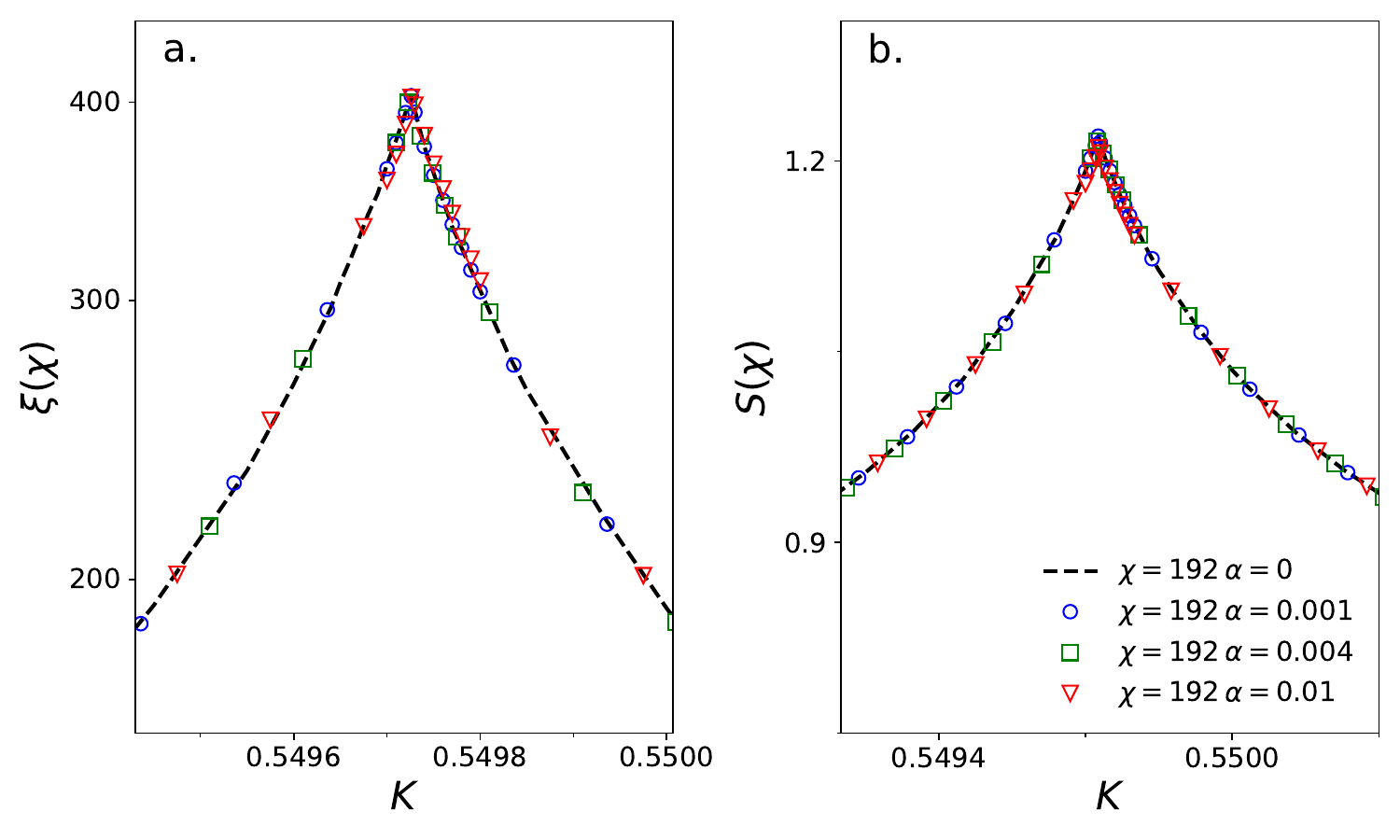}
\caption{\label{fig:3X} a. Correlation length $\xi(\chi)$ and b. entanglement entropy $S(\chi)$ as a function of $K$ across the critical point for the $X$ type weak measurement calculated for the measurement outcome $\left(\downarrow \downarrow \right)$. }
\end{figure}
\end{center}
\begin{center}
\begin{figure*}[ht]
\includegraphics[scale=0.35]{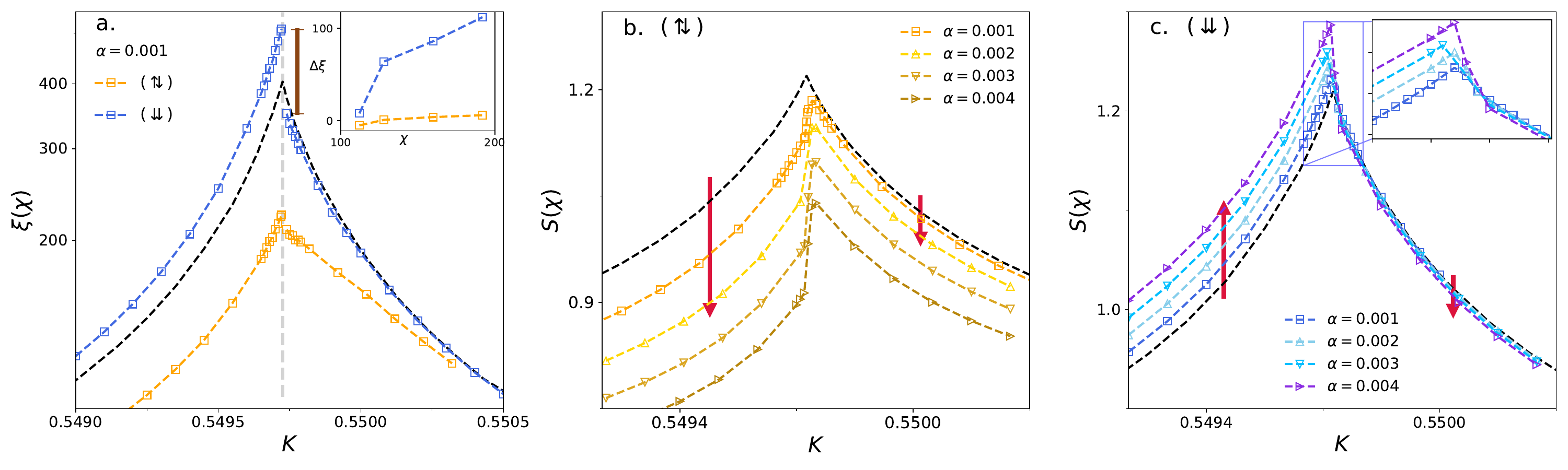}
\caption{\label{fig:3} a. Correlation length $\xi(\chi)$ as a function of $K$ at $\alpha=0.001$ following $Z$ type weak measurements. The gray vertical dashed line is drawn at the pseudo critical coupling $K_c(\chi)\approx 0.549726$ at $\chi=192$. The black dashed curve is the $\xi(\chi)$ of the critical ground state. The blue dashed curve with open squares represent the $\xi(\chi)$ of the measurement outcome $\left(\downarrow \downarrow\right)$. The $\xi(\chi)$ increases significantly when $K<K_c(\chi)$ and decreases weakly when $K>K_c(\chi)$.  The orange dashed curve with open squares represent the $\xi(\chi)$ of the measurement outcome $\left(\uparrow \downarrow\right)$ and shows overall reduction of $\xi(\chi)$.  The gap at the $K_c(\chi)$ is denoted by $\Delta \xi $. The inset shows the variation of  $\Delta \xi $ against the bond dimension for the measurement outcome $\left(\uparrow \downarrow\right)$-orange curve and $\left(\downarrow \downarrow\right)$-blue curve. The plots b. and c. shows the bipartite entanglement entropy $S(\chi)$ as a function of $K$ at $\alpha=0.001, 0.002, 0.003, 0.004$ for measurement outcome $\left(\uparrow \downarrow\right)$ and $\left(\downarrow \downarrow\right)$ respectively. In plot b. $S$ decreases when increasing the $\alpha$ and the rate of decrement is higher when $K<K_c(\chi)$. Note the direction and the length of red arrows drawn proportional to the rate of decrement. In plot c. $S$ increases when increasing $\alpha$ when $K<K_c(\chi)$(note the red up arrow) and $S$ decreases weakly when increasing $\alpha$ when $K>K_c(\chi)$ (note the red down arrow).    }
\end{figure*}
\end{center}

The measurements alter the entanglement structure of the post-measurement states at non zero values of $\alpha$. 
Thus, the physical observables of post-measurement states would deviate from their trivial reference values at $\alpha=0$ as increasing the measurement strength $\alpha$. We define weak measurements for smaller values $\alpha<< 1$ and find drastically different responses for the $X$ and $Z$ type of measurement outcomes. It should be noted that the values of $u$ and $\alpha$ are represented in unit of $\pi$ though out this paper. 

The $X$ type weak measurement results negligible alterations in correlation length and entanglement entropy for weak measurement strength. Especially, we consider $\alpha \sim \mathcal{O}(10^{-3})$ or smaller. In figure \ref{fig:3X}, it is plotted the variation of correlation length $\xi(\chi)$ and entanglement entropy $S(\chi)$ at $\chi=192$ for different values of $\alpha$ for the measurement outcome $\left(\downarrow \downarrow \right)$. Note the deviation of the data points from the $\alpha=0$ curve (dashed line) is negligible up to $\alpha=0.01$. We observe that the measurement outcomes $\left(\uparrow \downarrow \right)$ and $\left(\uparrow \uparrow \right)$ also show similar behaviour for smaller $\alpha$.  Thus, we verify that the $S(\chi)$ has logarithmic scaling such that $S(\chi)=\frac{c}{6}\ln(\xi(\chi))$ with estimated effective central charge $c_{eff}\approx 1$ for $\left(\downarrow \downarrow \right)$, $\left(\uparrow \uparrow \right)$ and $\left(\uparrow \downarrow \right)$ measurement outcomes studied in this work. Further, the order parameter also shows negligible variation from that of  $\alpha=0$.  Hence, we may conclude that post-measurement states resulted from $X$ type weak measurements would preserve the characteristic features of DQCP. On the other hand, the critical ground state shows extreme sensitivity to the $Z$ type weak measurements. Thus, in proceeding our work is mainly focus on the $Z$ type weak measurements. 

Fig.\ref{fig:3}a shows the $\xi(\chi)$ versus $K$ for $Z$ type measurements at $\alpha=0.001$ and $u=1/10$. The gray dashed vertical line represents approximately the pseudo critical coupling $K_c(\chi)\approx 0.549726$ at $\chi=192$. 
In the measurement outcome $\left(\uparrow \downarrow\right)$, strong reduction of $\xi(\chi)$ is observed when  $\alpha=0.001$. See the orange dashed line with squares in Fig. \ref{fig:3}a.
For the measurement outcome $\left(\downarrow \downarrow\right)$, $\xi(\chi)$ is large extending more than 400 lattice  spacings while showing strong phase dependence. See the blue dashed line with squares in Fig.\ref{fig:3}a. In the zFM phase where $K<K_c(\chi)$,  $\xi(\chi)$ increases beyond the black dashed line representing the  $\xi(\chi)$ at $\alpha=0$. In the VBS phase where $K>K_c(\chi)$, $\xi(\chi)$ decreases slightly below $\alpha=0$. Particularly, there is a discontinuity developing at  $K_c(\chi)$ in both measurement outcomes, but prominently in the measurement outcome $\left(\downarrow \downarrow\right)$.  Therefore, when reaching the $K_c(\chi)$ from high $K$ and low $K$ sides, the $\xi(\chi)$ has different values at the $K_c(\chi)$. 
This is a strong signature of a weak first order transition \cite{gangat2024prb}.
Further, it would leads to phase coexisting regions which would preserve in the thermodynamic limit. A discussion on phase coexistence of post-measurement states with numerical results is presented in the Appendix \ref{app:orderp}.

\begin{center}
\begin{figure}[ht]
\includegraphics[scale=0.32]{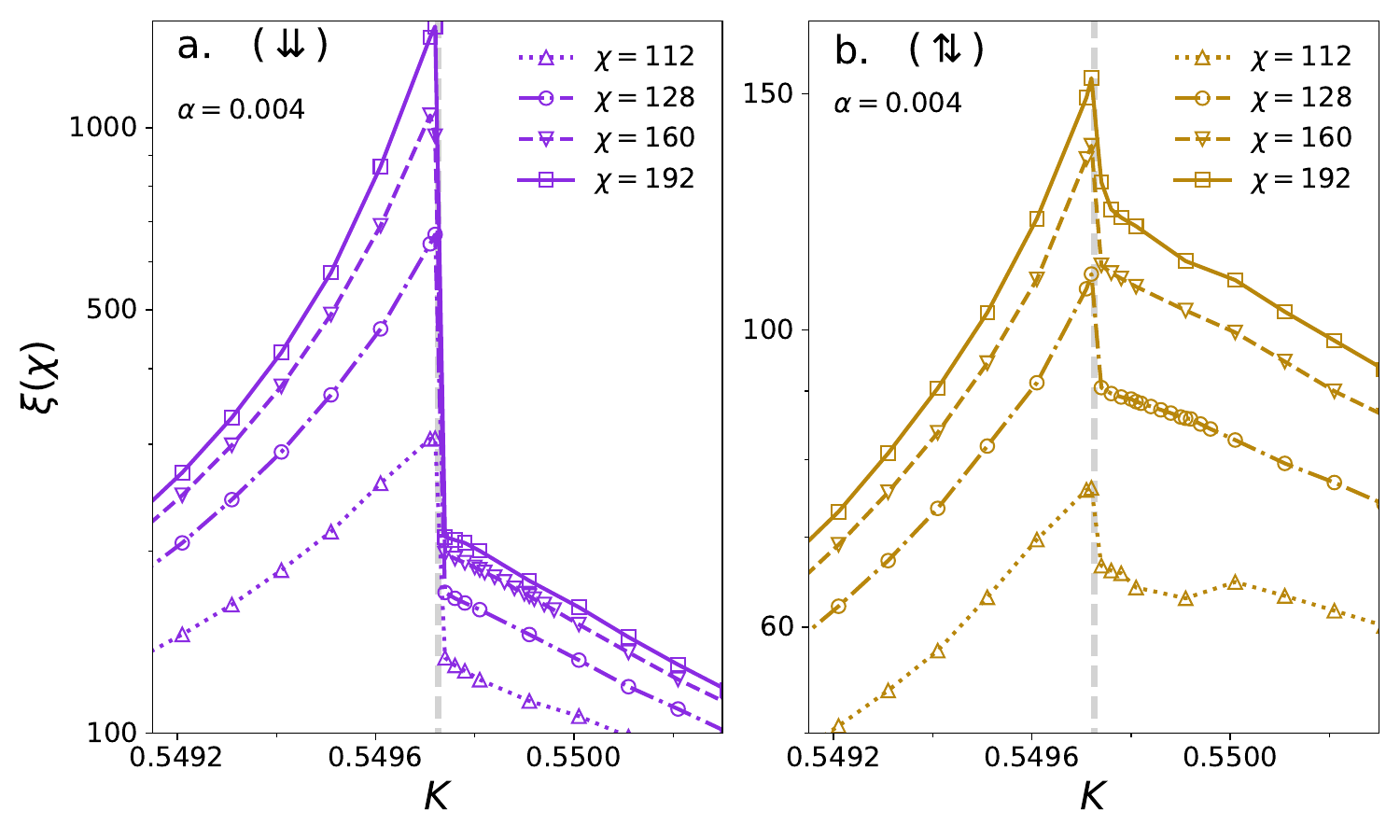}
\caption{\label{fig:32} Correlation length $\xi(\chi)$ as a function of $K$ across the critical coupling (dashed vertical line) for the the measurement outcome a. $\left(\downarrow \downarrow \right)$ and b.$\left(\uparrow \downarrow \right)$. }
\end{figure}
\end{center}
The gap of $\xi(\chi)$  at $K_c(\chi)$, $\Delta \xi$ increases when increasing the $\chi$ for both trajectories of measurement outcomes showing a considerable higher rate in measurement outcome $\left(\downarrow \downarrow\right)$. See the inset of Fig.\ref{fig:3}a. Further, it is also confirmed that the $\Delta \xi$ increases when increasing the maximum strength of weak measurement $\alpha$. In figure \ref{fig:32}, it is plotted the $\xi(\chi)$ as a function of $K$ across the critical coupling at $\alpha=0.004$. Note the gap at the critical coupling $\Delta \xi$ increases rapidly with $\chi$ for the measurement outcome $\left(\downarrow \downarrow\right)$. It is also noted a overall decrease of the $\xi(\chi)$ when increasing $\alpha$ for the measurement outcome $\left(\uparrow \downarrow\right)$ while $\Delta \xi$ increases weakly with the $\chi$. Thus, it confirms that a finite gap $\Delta \xi$ would exist in the limit  $\chi \rightarrow \infty$ and  the weak first order nature of the phase boundary could be more prominent in the thermodynamic limit.

To gain more insight on weak measurement with $Z$ type measurement operators, we calculate the spatially averaged von Neumann bipartite  entanglement entropy($S$) of the post measurement states. In the $\ket{\psi_{gs}}$, $S$ increases initially with $K$ and  shows a peak at the vicinity of the pseudo critical coupling $K_c(\chi)$ and decreases when further increasing $K$. See the black dashed line in Fig.\ref{fig:3}b and c.  Figure \ref{fig:3}b shows the  $S$ versus $K$ for different values of $\alpha$ for the post measurement state  $\ket{\psi_{(\uparrow \downarrow)}^Z}$. The $S$ decreases as increasing $\alpha$, with higher rate when $K<K_c(\chi)$  compared to $K>K_c(\chi)$. The bipartite entanglement entropy  represent non-local correlations including both long-range and the short-range correlations in the system. The reduction of entanglement entropy could lead to convert the long range correlations to the short-range once. Thus, it would reduce the longest possible correlations in the system and hence reduce the correlation length as calculated via the transfer matrix method (orange dashed line in the Fig. \ref{fig:3}a).  It is also evident from the Fig. \ref{fig:3}b, that the restructuring of the entanglement is not monotonically varying across the phase boundary, hence develop a gap in the entanglement entropy and also in the correlation length at the vicinity of the pseudo critical coupling. 

An anomalous behaviour is observed in the measurement outcome $\left(\downarrow \downarrow \right)$. see Fig.\ref{fig:3}c. The $S$ strongly increases as increasing $\alpha$ when $K<K_c(\chi)$ and weakly decreases when $K>K_c(\chi)$ in the vicinity of the pseudo critical point. The above highly asymmetric response of the entanglement across the phase boundary can be identified as an anomalous behaviour. The benchmarking calculations conducted on the transverse field Ising model suggests that the  response of $S$ due to measurement varies monotonically across the phase boundary in conventional critical points.  The increment of $S$ when $K<K_c(\chi)$ could lead to restructure the short-range correlations to long-range correlations which results overall increment in the $\xi$ in the system  as shown by the blue dashed line in Fig.\ref{fig:3}a.
We confirm that the  $\xi$ always follows the above trend of asymmetric restructuring of the $S$  as increasing the strength of the weak measurement $\alpha$. 
Thus, the asymmetric profile of correlation length $\xi$ shown in \ref{fig:3}a could be a direct manifestation of the developed asymmetric entanglement structure under the weak measurement.  

In our formalism the measurement operators are defined with two independent parameters $u$ and $\alpha$.   It is verified that the parameter $u$ solely determine the probability of the post measurement state in the limit of weak measurement $\alpha<<1$. 
The probability of obtaining the measurement outcome $m$ can be given by $P_m=\bra{\psi_m}\ket{\psi_m}$. For $\alpha=0$, we can express the $P_m$ in simple analytical form in the basis of two spins in the measurement basis. Initiating all the ancilla spins in the $\ket{\downarrow} $ state, it can be derived that, $P_{\left(\downarrow \downarrow \right)}=\cos^4(u)$, $P_{\left(\uparrow \downarrow \right)}=P_{\left(\downarrow \uparrow \right)}=\sin^2(u) \cos^2(u)$, and $P_{\left(\uparrow \uparrow \right)}=\sin^4(u)$. Thus, the $\left(\downarrow \downarrow \right)$  is the most probable measurement outcome for small $u$ whereas the $\left(\uparrow \uparrow \right)$  is the least probable measurement outcome for small $u$.  It is numerically verified that the above measurement probabilities constitute a good approximation for Born-rule probabilities  when $\alpha \leq 0.01$. See the Appendix \ref{app:born_rule}.

\begin{center}
\begin{figure}[ht]
\includegraphics[scale=0.35]{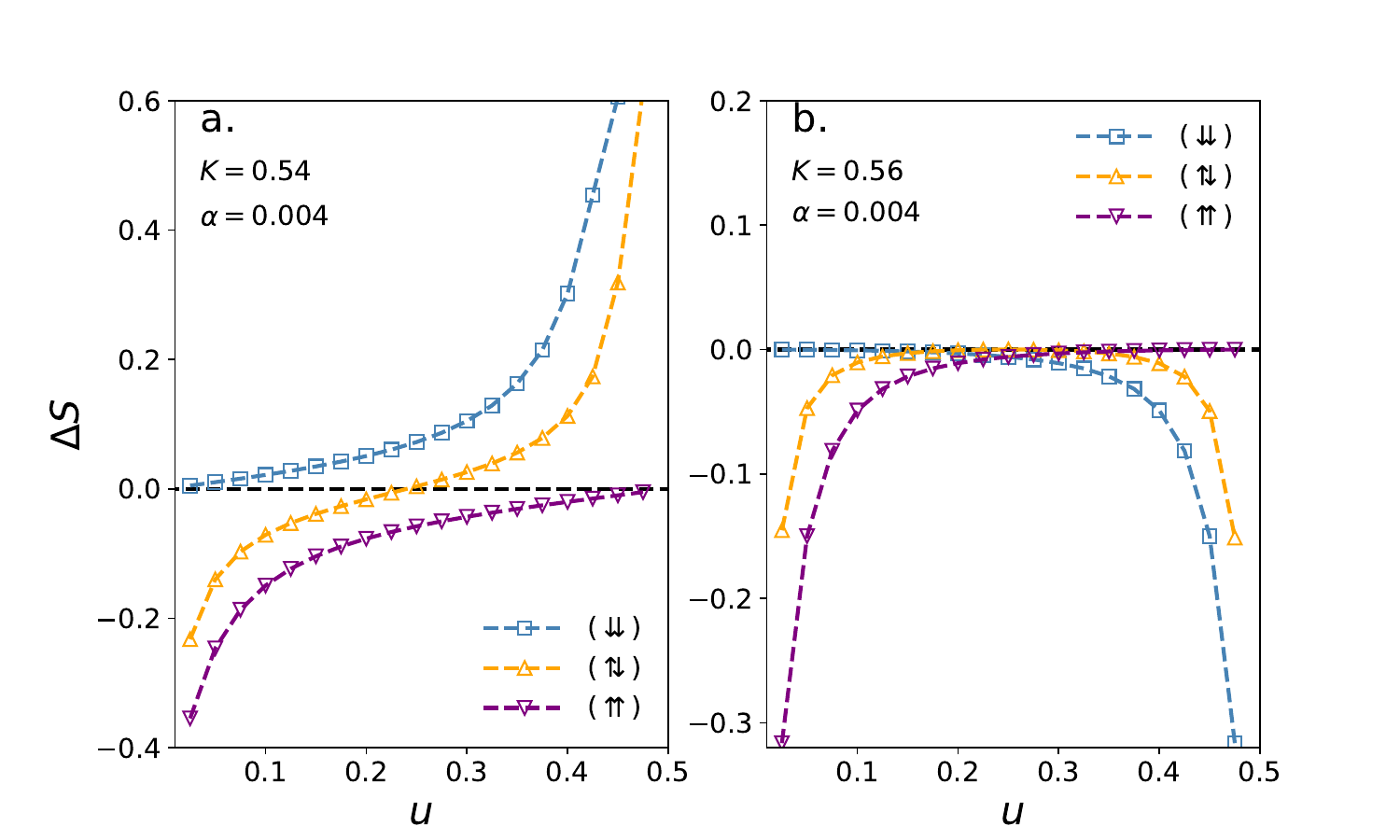}
\caption{\label{fig:6} Shift of entanglement entropy $\Delta S$ versus the unitary parameter $u$.  a.) at $K=0.54$  ($<K_c$) and b.) at $K=0.56$ ($>K_c$),  for measurement outcomes $\left(\downarrow \downarrow \right)$, $\left(\uparrow \downarrow \right)$ and $\left(\uparrow \uparrow \right)$ when $\alpha=0.004$.  }
\end{figure}
\end{center}
The $\Delta S$ is the shift of entanglement entropy of the post measurement state $S_m$ from that of the initial ground state $S_{gs}$, $\Delta S= S_m-S_{gs}$.  In Fig.\ref{fig:6}, it is depicted the $\Delta S$ of post-measurement states against the unitary parameter $u$ in the range $(0 - 1/2$) at  $\alpha=0.004$ for $Z$ type weak measurements. 
Here, $\Delta S$ of the measurement outcomes $\left(\downarrow \downarrow \right)$, $\left(\uparrow \downarrow \right)$, and $\left(\uparrow \uparrow \right)$ are calculated at $K=0.54$(Fig.\ref{fig:6}a.) and  $K=0.56$(Fig.\ref{fig:6}b.). 

The  measurement outcome $\left(\uparrow \uparrow \right)$ measuring each ancilla spin state flipped,  shows the lowest negative shift of $\Delta S$ at $K=0.54$. See the purple dashed line with down triangles in Fig.\ref{fig:6}. At $K=0.56$, the $\Delta S$ also stays the lowest when $u<1/4$ and reaches to zero when $u>1/4$.  The strength of weak measurement in a spin flip $\lambda^{flip} \sim \alpha \cos(u)$. Thus, the maximum of $\lambda^{flip}_{max}\sim \alpha$ occurs when $u\rightarrow 0$ and decreases when increasing $u$. The numerically calculated $\Delta S$ shows maximum reduction at $u=0$ and $\Delta S$ increases slowly when the $\lambda^{flip}$ decreases with increasing $u$. 

The  measurement outcome $\left(\downarrow \downarrow \right)$ measuring no change in the  initialized ancilla spin states,  shows $\Delta S >0$ at $K=0.54$ whereas $\Delta S <0$ at $K=0.56$ for the whole range of $u$.  The strength of weak measurement in ancilla spin measured in the initialized state is $\lambda^{ini} \sim \alpha \sin(u)$. Thus, the $\lambda^{ini} \rightarrow 0$  when $u\rightarrow 0$ and the maximum occurs at $u=1/2$. Thus, there is no change of $\Delta S$ when $u\rightarrow 0$ since the strength of weak measurement is negligible. The $\lambda^{ini}$ increases when increasing $u$ so that $\Delta S$ increases positively which is the anomalous behaviour observed.
However, when $K=0.56$,  $\Delta S$ decreases negatively as increasing $u$. 
 This is highlighting main result of this work showing the prominent asymmetric restructuring of entanglement entropy. This confirms the asymmetric entanglement restructuring in the two phases across the phase boundary is consistent for broader range of measurement parameters $u$ and $\alpha$. 

The $\Delta S$ of the measurement outcome $\left(\uparrow \downarrow \right)$ shows a mixed effect in both phases. The $\Delta S$ follows the trend of the outcome $\left(\uparrow \uparrow \right)$ when $u<1/4$ and that of $\left(\downarrow \downarrow \right)$ when $u>1/4$ since the strength of weak measurement in each outcome is dominating in this region. See the orange dashed line with up triangles in Fig.\ref{fig:6}. 

In the context of weak measurement described above, measuring the outcome $\left(\downarrow \downarrow\right)$  has  highest probability when $u \rightarrow 0$. However, it is not possible to observe measurement induced effects in the post measurement state since the strength of the measurement is negligible. The outcome $\left(\uparrow \uparrow\right)$ has highest effect due to the weak measurement when $u\rightarrow 0$, but the those outcomes are highly unlikely.  
Thus, it may be necessary to postselect the desired outcomes at sufficiently larger $u$ to capture the measurement induced effects in the post measurement states. 
 
\section{\label{con:disc} Discussions and Conclusions}
In this paper, we study the effect of weak measurement on deconfined quantum critical phases transition. The critical ground state is probed by coupling to an ancilla and later measuring the ancilla spins projectively.  In our measurement protocol, critical system-ancilla coupling is  modelled using unitary interaction which depends on two parameters representing  unitary operations ($u$) and non unitary action($\alpha$)  restructuring entanglement across the critical system. When $\alpha=0$, the derived effective measurement operators act as local unitary operators and observe trivial variations of the order parameters which depends on the measurement outcome. We find that the probability of measurement outcomes depends solely on the unitary parameter $u$ when $\alpha=0$ and  order parameter amplitudes of each phases are independent of  $u$. We numerically verify that the simple analytical form of measurement probability derived at $\alpha=0$ is in a good approximation in the weak measurement limit defined as $\alpha <<1$.   

We find that in the weak measurement limit, the bipartite entanglement entropy($S$) of the post-measurement states under $X$ type of measurement follow logarithmic scaling near the critical point yielding the effective central charge $c\approx1$ and hence preserve the salient features of DQCP. We verify that the  above effect is consistent for $\alpha<0.01$ and highly deviate when increasing $\alpha >0.01$. The $S$ and $\xi$ shows asymmetric development at higher values $\alpha$ with growing coexisting region near the phase boundary.  

It is mainly studied the effect of $Z$ type measurement on the DQCP which shows an anomalous entanglement restructuring  across the phase boundary. Especially, highly asymmetric entanglement entropy is developed between the $K<K_c$ and $K>K_c$ regions. We numerically find that the asymmetry of $\Delta S$ is anomalous in the measurement outcome $\left(\downarrow \downarrow \right)$.  The $\Delta S$ is positively increasing when $K<K_c$ and negatively decreasing when  $K>K_c$. It is found that the above changes in $S$ indeed related to the restructuring of the local correlations in the system. We observe that the developed asymmetry of the $\xi(\chi)$ across the phase boundary get manifested at the vicinity of the  critical point resulting finite gap $\Delta \xi(\chi)$ at the critical point which would persist in the thermodynamic limit as $\chi \rightarrow \infty$. Further, it is also observed developing strong coexisting phases in the post measurement states which would persist as $\chi \rightarrow \infty$ (Appendix \ref{app:orderp}). The coexistence is prominent when the entropy is decreasing in the post measurement states especially in the $K>K_c$ region with measurement outcomes $\left(\uparrow \downarrow \right)$. The above two numerical evidences presented in this work following extensive numerical calculations strongly support the weakly first order nature of the phase boundary of post measurement states. 

The observed prominent enhancement of  $S$ near the criticality when $K<K_c$ could be due to the long range(next nearest neighbour) interactions of the Hamiltonian. The benchmarking calculations conducted on 1D quantum Ising chain with only short range(nearest neighbour) interactions ($g$) reveals the enhancement of entanglement near the critical point($g_c$) in the magnetic phase is not visible for the outcome $\left(\downarrow \downarrow \right)$. A significant enhancement of $S$ is observed deep in the magnetic phase. On the other hand the $S$ decreases significantly in the disordered phases when $g>g_c$ compared to the VBS phase. In two-dimensional(2D) systems the increased lateral connectivity would results extensive growth of entanglements beyond the 1D systems with long range interactions. Thus, the enhancement of entanglement near the criticality  in the magnetic phase would be significant in 2D compared to the 1D.     However, it may require comparatively larger measurement strength to suppress the entanglement in VBS phase in 2D. Thus, we would observe a  higher degree of asymmetry in the entanglement entropy at the DQCP in 2D for the measurement outcome $\left(\downarrow \downarrow \right)$.

In our formalism, there is a competing effect on probability of measurement and the strength of weak measurement ($\lambda$), which depend on the parameters $u$ and $\alpha$ defining the unitary interactions between critical system and ancilla coupling. Thus, it may require to postselect specific outcomes to observe the measurement induced effects. 
Thus, we believe that with the recent proposals to experiment DQCP in novel quantum simulators \cite{lee2023prl}, the measurement induced effects can also be tested by tuning the interaction with the measurement apparatus.

\begin{acknowledgments}
I would like to acknowledge Teemu Ojanen and Bruno Uchoa for assistance provided and useful discussions. I would like to acknowledge University of Ruhuna Sri Lanka, where I was affiliated during the initial part of this independent study. Finally, I would also like to acknowledge the anonymous referees who provided constructive feedbacks on this work. 

\end{acknowledgments}

\appendix

\section{\label{app:orderp} Order parameters }

The quantum phases are analysed by calculating the order parameters of valance bond solid $\langle O_{VBS}\rangle$, ferromagnetic $\langle O_{zFM}\rangle$ and anti-ferromagnetic $\langle O_{zAFM}\rangle$ phases as follows, 
\begin{eqnarray}
\langle O_{VBS}\rangle &=& \frac{1}{2}\sum_{j\in \{1,2\}} \langle e^{i\pi j}\left(\vec{\sigma}_{j-1}.\vec{\sigma}_j-\vec{\sigma}_j.\vec{\sigma}_{j+1}  \right)\rangle \nonumber \\
\langle O_{zFM}\rangle &=& \frac{1}{2}\sum_{j\in \{1,2\}} \langle \sigma^z_j \rangle  \\
\langle O_{zAFM}\rangle &=& \frac{1}{2}\sum_{j\in \{1,2\}} \langle e^{i\pi j}\sigma^z_j \rangle \nonumber
\label{eq:odp}
\end{eqnarray}
The above averages are calculated over a unit cell. 

The order parameters of the post-measurement states at $\alpha=0$ also vanishes simultaneously when the exchange interaction $K$ reaching the critical point.  
However, the MPS study  gives a first order phase transition at finite MPS bond dimension $\chi$. Ref. \cite{roberts2019prb} shows that there exists a continuous transition in the limit of $\chi \rightarrow \infty $ using finite size scaling with bond dimension for the ground state of the Hamiltonian (\ref{Hqs}). Following the same prescription, we find that the order parameters of the post-measurement states also show characteristic critical behaviour of the DQCP when $\alpha=0$. 

\begin{center}
\begin{figure}[ht]
\includegraphics[scale=0.42]{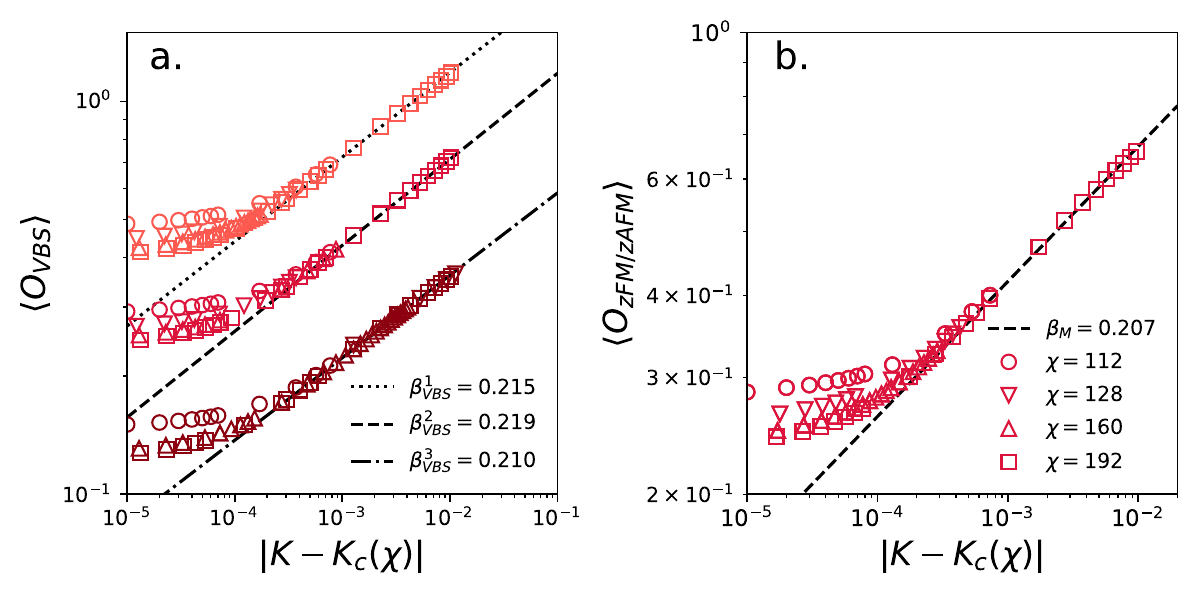}
\caption{\label{fig:ap1} Numerically calculated order parameters as a function of $|K-K_c(\chi)|$. In the left panel (a.) the 3 sets of curves in three different colors are for post measurement states $\ket{\psi_{(\uparrow \downarrow)}^X}$-(orange symbols), $\ket{\psi_{(\downarrow \downarrow)}^{X/Z}}$-(red symbols) and $\ket{\psi_{(\uparrow \downarrow)}^Z}$-(brown symbols) from top to bottom. In the right panel(b.), the $\langle O_{zFM} \rangle$ and $\langle O_{zAFM} \rangle$ of different post-measurement states coincide. Note in the regions $K<K_c$ and $K>K_c$, the order parameters follows  approximately the same power-law exponent $\beta$ for different $\chi$. }
\end{figure}
\end{center}

Fig. \ref{fig:ap1} shows the order parameters as a function of $|K-K_c(\chi)|$ for different $\chi$. There is a region closer to the transition which follows the power-law  scaling such that $\langle O_x \rangle \sim |K-K_c(\chi)|^{\beta} $  and extends towards the critical point when increasing $\chi$. 
The order parameter exponents, $\beta$'s of VBS phases for the 3 different set of curves plotted in Fig. \ref{fig:ap1} (left panel) has approximately similar values. The exponents $\beta_1$, $\beta_2$, and $\beta_3$ are correspond to the post measurement states $\ket{\psi_{(\uparrow \downarrow)}^X}$, $\ket{\psi_{(\downarrow \downarrow)}^{X/Z}}$ and $\ket{\psi_{(\uparrow \downarrow)}^Z}$ respectively. In the magnetic phases $K<K_c(\chi)$, the $\langle O_{zFM} \rangle$ and $\langle O_{zAFM} \rangle$ of different post-measurement states coincide as shown in Fig.\ref{fig:2} in the main text. The Figure \ref{fig:ap1}  (right panel) shows the order parameters in the logarithmic scale showing the power-law fitting. The estimated order parameter exponent for the magnetic phases ($\beta_M$) is approximately equal to that of VBS phases. 

\begin{center}
\begin{figure}[ht]
\includegraphics[scale=0.33]{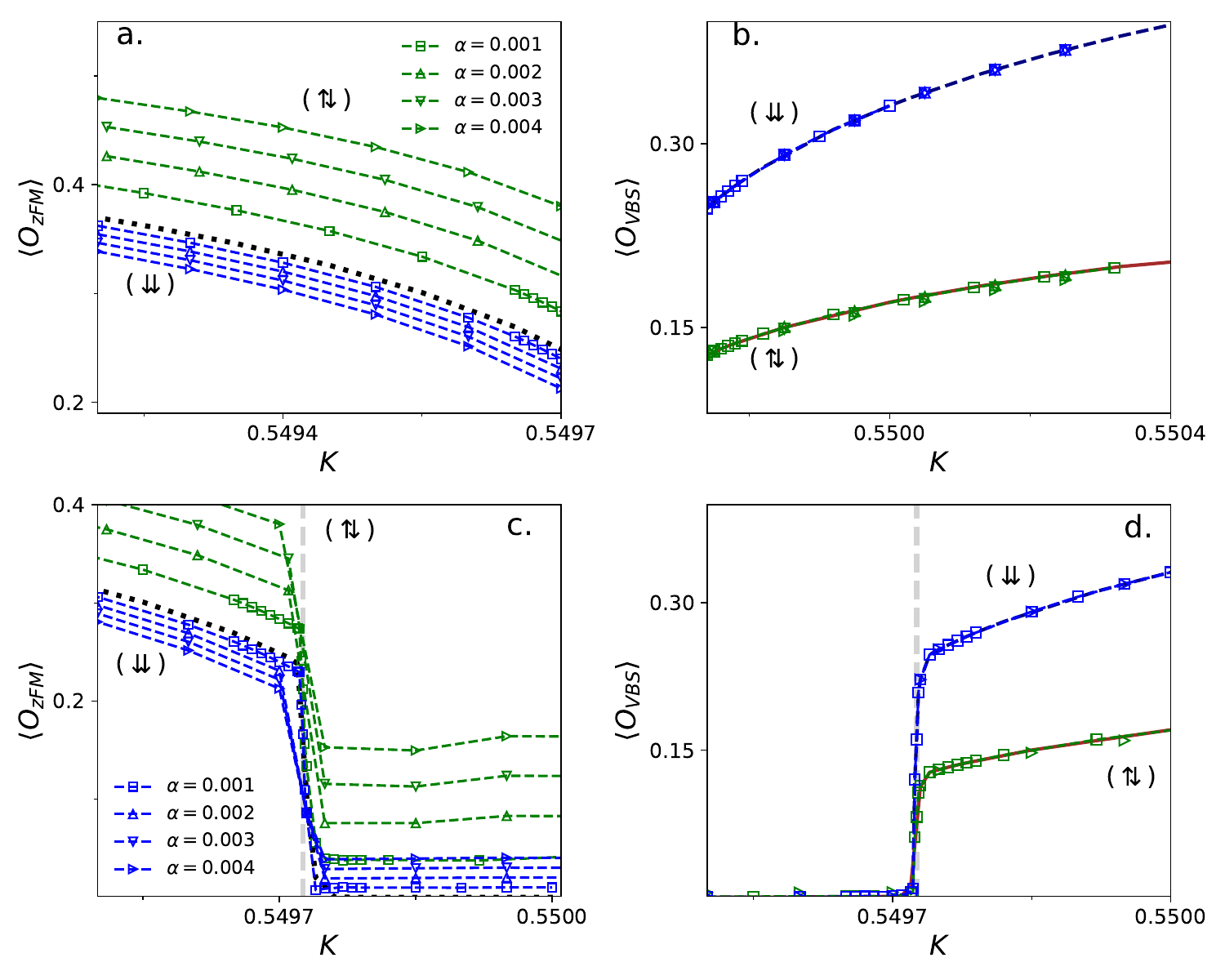}
\caption{\label{fig:opweak} Order parameters, Fig. a. $\langle O_{zFM} \rangle$ phase  and Fig. b. $\langle O_{VBS} \rangle$ phase calculated for $Z$ type weak measurements. The Fig. c. and d. show the phase coexistence calculated at the maximum bond dimension ($\chi=192$) employed in the VUMPS calculation. The gray dashed vertical line represent the approximate critical coupling of the critical ground state evaluated at $\chi=192$.  }
\end{figure}
\end{center}

When the weak measurement is carried out, the magnitude of the order parameters varies from the above $\alpha=0$ values of each measurement outcomes. In Fig. \ref{fig:opweak}a. the dotted black line shows the $\langle O_{zFM}\rangle$ for all form of studied measurement outcomes when $\alpha=0$.  The dashed black line and the red solid lines in the Fig. \ref{fig:opweak}b. represent the $\langle O_{VBS}\rangle$ for the measurement outcomes $\left(\downarrow \downarrow \right)$ and $\left(\uparrow \downarrow \right)$ respectively when $\alpha=0$. 

The response of the order parameters for $Z$ type weak measurement  is represented in Fig \ref{fig:opweak}. The green dashed lines with symbols represent the measurement outcome $\left(\uparrow \downarrow \right)$ and the blue dashed lines with symbols represent the measurement outcome $\left(\downarrow \downarrow \right)$ . The system disentangle as a result of reduced entanglement entropy in the outcome $\left(\uparrow \downarrow \right)$. Thus, the $\langle O_{zFM}\rangle$ increases as increasing $\alpha$( green  dashed lines in Fig. \ref{fig:opweak}a.). However, the shift of the $\langle O_{VBS}\rangle$ is negligible despite of the reduced entanglement entropy. It is evidenced that later trend above is due to the developing zFM phase in the $K>K_c(\chi)$ region coexisting with the VBS phase(see Fig.\ref{fig:opweak}c.). The phase coexistence is prominent and the  $\langle O_{zFM}\rangle$ is comparable to the $\langle O_{VBS}\rangle$ at $\alpha=0.004$ when $K>K_c(\chi)$.  

The $S$ of the measurement outcome $\left(\downarrow \downarrow \right)$ increases with the weak measurement strength $\alpha$ as described in the main text. 
Thus, the $\langle O_{zFM}\rangle$ decreases due to the increased entanglement when $K<K_c(\chi)$ (blue dashed lines with symbols in Fig. \ref{fig:opweak}a.).  The development of the coexisting zFM phase is also observed  in the region $K>K_c{\chi}$ with markedly lower $\langle O_{zFM}\rangle$ compared to the measurement outcome $\left(\downarrow \uparrow \right)$. This could be due to the weak response of the entanglement entropy  to $\alpha$ in the $K>K_c(\chi)$ regime.

The phase coexistence is an another evidence of the first order nature of the phase transition in the post measurement states. A weakly first order phase transition is characterised by the existence of a narrow coexisting region \cite{gangat2024prb}.  The Fig.  \ref{fig:opweak}c. and d. shows the developed coexisting regions into the VBS and zFM phases calculated  at $\chi=192$.  The $K_c$(gray vertical dashed line) is the approximate critical coupling of the intrinsic ground state at $\chi=192$. It is observed developing zFM phase in the $K>K_c$ region when increasing the $\alpha$ and extended deep in to the VBS phase (See Fig. \ref{fig:opweak} c.).  The coexistence of the VBS phase with the zFM phase in the $K<K_c$ region is negligible as shown in Fig. \ref{fig:opweak}d. Thus, the VBS phase shows robust behaviour subjected to the Z type weak measurement. Thus, it is evidenced that the entanglement restructuring is prominently alter the zFM ordering of the system. The $\langle O_{zFM}\rangle$ decreases slowly when increasing $K$ beyond the $K_c$. It is observed that the developed coexistence region in the post measurement states has a very weak dependence on $\chi$ unlike the behaviour of the critical ground sate described in the Fig.\ref{fig:ap1}.  Thus, we believe that the coexistence is strong and would persist to the thermodynamic limit as $\chi \rightarrow \infty$. However locating the boundaries of the coexisting regions and detail characterization of the coexisting phases is beyond the limits of the numerical calculations employed.

\section{\label{app:born_rule} Measurement outcomes and the Born rule probability of post measurement states. }
In this work, we study the post measurement states in a broader range of unitary parameters $u$. In most of the studies reported recently have only considered the scenario when $u=1/4$ where the measurement induced backaction is minimum. However, we have observed in this work, that studying the measurement induced effect in a broader range of unitary coupling sheds lights on unveiling important aspects related to the measurement induced entanglement(MIE). As shown by the Fig.\ref{fig:6} in the main text, a descriptive understanding of the entanglement of the post measurement states can be obtained.
\begin{center}
\begin{figure}[ht]
\includegraphics[scale=0.5]{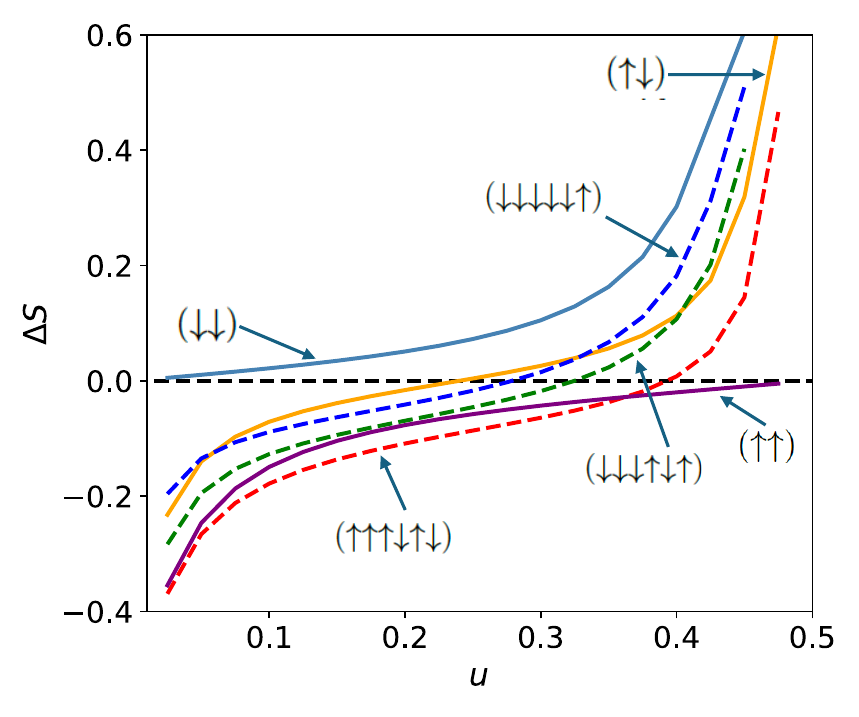}
\caption{\label{fig:9} Shift of entanglement entropy $\Delta S$ versus unitary parameter u for  different measurement outcomes having 6 spins in the measurement basis. }
\end{figure}
\end{center}
In the main text, the measurements are carried out periodically keeping two spin moments in the measurement basis within the uniform MPS representation of the ground sate wave function. This procedure can be extended to larger measurement basis to study more disordered like measurement outcomes. In Fig. \ref{fig:9}, it is  depicted the variation of $\Delta S$ for few post measurement states keeping 6 spin moments in the measurement basis.  The $\Delta S$ for uniform measurement outcomes: $\left(\downarrow \downarrow \downarrow \downarrow \downarrow \downarrow\right)$ and $\left(\uparrow \uparrow \uparrow \uparrow \uparrow \uparrow\right)$ and the outcome with alternative spin orientation $\left(\uparrow \downarrow \uparrow \downarrow \uparrow \downarrow \right)$ is same as the two basis measurement. The other measurement outcomes with mixed spin states: $\left(\downarrow \downarrow \downarrow \downarrow \downarrow \uparrow\right)$, $\left(\downarrow \downarrow \downarrow \uparrow \downarrow \uparrow\right)$, and  $\left(\uparrow \uparrow \uparrow \downarrow \uparrow \downarrow\right)$ are represented by the blue, green and red dashed lines respectively as marked in the Fig. \ref{fig:9}.

The strength of weak measurement for an ancilla spin been measured in $\bra{\downarrow}$ and  $\bra{\uparrow}$ states is $\alpha \sin(u)$ and $\alpha \cos(u)$ respectively.   Thus, when $u\rightarrow 0$, the measurement effect on the entanglement entropy vanishes for the uniform measurement outcomes $\left(\downarrow \downarrow \cdots \downarrow\right)$. In contrary, the measurement strength increases to its maximum value ($\alpha$) for measuring $\bra{\uparrow}$ state when $u\rightarrow 0$.
Thus, the effect of the $\bra{\uparrow}$ state dominate in the entanglement entropy when $u\rightarrow 0$ in a measurement outcome with a single  $\bra{\uparrow}$ state with many $\bra{\downarrow}$ state. Similarly, the effect of $\bra{\downarrow}$ state dominates when $u\rightarrow 1/2$. This effect is reflected in the measurement outcome $\left(\downarrow \downarrow \downarrow \downarrow \downarrow \uparrow\right)$ depicted by the blue dashed line in Fig.\ref{fig:9}. The above argument can  be  applied to the all the other measurement outcomes with mixed spin states as shown by the green and red dashed lines in the Fig.\ref{fig:9}. The all measurement outcomes with mixed spin states would qualitatively follow the physics of the measurement outcome with alternative spin orientations $ \left(\uparrow \downarrow \right)$ in the weak measurement limit defined as $\alpha << 1$. 

\subsection*{Born-rule probability of post measurement states.}
\begin{center}
\begin{figure}[ht]
\includegraphics[scale=0.43]{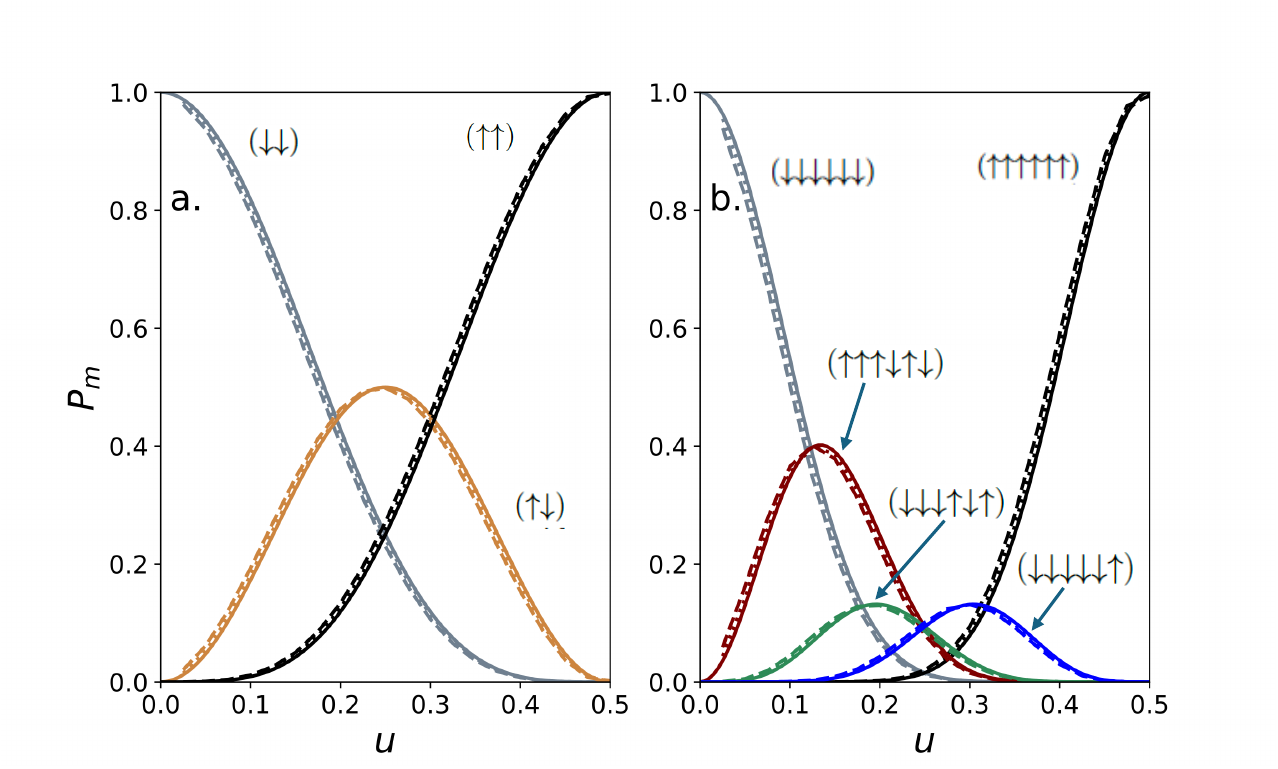}
\caption{\label{fig:10} Born-rule probabilities for measurement outcomes with 2 spin  moments in the measurement basis (a.) and 6 spin moments in the measurement basis (b.). The solid lines are the theoretical probabilities when $\alpha=0$. The dotted lines and the dashed lines are from the numerical calculation when $\alpha=0.004$ and $\alpha=0.01$ respectively.}
\end{figure}
\end{center}
The Born-rule probability of a measurement outcome $m$ can be given by, $P_m=\bra{\psi_m}\ket{\psi_m}$. It is straight forward to note  when $\alpha=0$, the contribution to the normalization constant measuring a single spin in $\bra{\downarrow}$ state is $\cos(u)$ and $\bra{\uparrow}$ state is $\sin(u)$. Thus, the contribution to the probability of the measurement outcome is $\cos^2(u)$ and $\sin^2(u)$. Now we can simply write the Born-rule probability of a unit cell correspond to the periodic measurement protocol discussed in the main text. For the case where two spin moments in the measurement basis, there are 3 distinct measurement outcomes as discussed in the main text. Thus, the probability of those outcomes can be write as, $P_{\left(\downarrow \downarrow\right)}=\cos^4(u)$, $P_{\left(\uparrow \uparrow\right)}=\sin^4(u)$ and $P_{\left(\uparrow \downarrow\right)}=2\cos^2(u)\sin^2(u)$. The factor 2 in the outcome $\left(\uparrow \downarrow\right)$  is due to the 2 possible degenerate states. 
Similarly, the probability of the outcomes in 6 basis measurement can be written as a product of $\sin^2(u)$ and $\cos^2(u)$ terms with a multiplication factor of number of degenerate trajectories with similar spin arrangements. For example, the probability of some of the measurement outcomes can be written as, $P_{\left(\downarrow \downarrow \downarrow \downarrow \downarrow \downarrow \right)}=\cos^{12}(u)$, $P_{\left(\downarrow \downarrow \downarrow \downarrow \downarrow \uparrow \right)}=6 \cos^{10}(u)\sin^2(u)$, $P_{\left(\downarrow \downarrow \downarrow \uparrow \downarrow \uparrow \right)}=6 \cos^{8}(u)\sin^4(u)$, and $P_{\left(\uparrow \uparrow \uparrow \downarrow \uparrow \downarrow \right)}=6 \cos^{4}(u)\sin^8(u)$.  It must be noted that the sum of the probabilities of all the possible measurement outcomes satisfy the probability conservation, $\sum_{\langle m \rangle } P_m =1$. In general for $n$ spin moments in the measurement basis, the probabilities of all the possible measurement outcomes can be given by the Binomial expansion such that $\left(\cos^2(u)+\sin^2(u)\right)^n=1$. For example, for periodic measurement outcomes with $r$ number of $\uparrow$ spins and $(n-r)$ number of $\downarrow$ spins, the total probability of such  outcomes can be given by $P_{(n,r)}= \leftindex^n{C}_r \cos^{2(n-r)}(u)\sin^{2r}(u)$. The Binomial coefficient $\leftindex^n{C}_r$ gives total degeneracy of periodic trajectories with $r$ number of $\uparrow$ spins.  We observe that the above theoretical Born-rule probabilities evaluated at $\alpha=0$ are in good approximation with the probabilities evaluated from the numerical calculation for  weak measurement limit $\alpha \leq 0.01$ as shown in the Fig. \ref{fig:10}.


\bibliography{DQCM_ref.bib}

\end{document}